\documentstyle[twocolumn,aps,epsf]{revtex}

\draft

\begin{document}
\title{Time-reversal violating rotation of polarization plane of light in
gas placed in electric field}
\author{ V. G. Baryshevsky\thanks{%
E-mail bar@inp.minsk.by}, D. N. Matsukevich\thanks{%
E-mail mats@inp.minsk.by}}
\address{Institute of Nuclear Problems, Belarusian State University \\
St. Bobryiskaya 11, 220050, Minsk, Republic of Belarus }
\date{\today}
\maketitle

\begin{abstract}
Rotation of polarization plane of light in gas placed in electric
field is considered. Different factors causing this phenomenon are
investigated. Angle of polarization plane rotation for transition $6S_{1/2}
\rightarrow 7S_{1/2}$ in cesium ($\lambda=539$nm) is estimated. The
possibility to observe this effect experimentally is discussed.
\end{abstract}


\narrowtext

\section{INTRODUCTION}

Violation of time reversal symmetry was observed in $K_0$ meson \cite%
{Cristenson,CPLEAR} and $B_0$ meson decay \cite{Unknown} and remains one of
the greatest unsolved problems in the elementary particle physics. A lot of
attempts have been undertaken to observe time-reversal violating phenomena
in different processes experimentally. However, those experiments have not
been successful. Among them there are, for example, measurements of electric
dipole moment (EDM) of neutrons \cite{Neutron}, atoms and molecules \cite%
{Tl-EDM,Cs-EDM,TlF-EDM}. No EDM was found but these experiments impose
strong restrictions on the theory. At EDM of heavy atoms search, in
particular, tight limits for parameters of electron-nucleon $P$-, $T$-
violating interactions and value of electron EDM \cite{Tl-EDM} are
set.

It is well known that essential progress in measurements of $P$-odd
interaction constants was achieved at study of the optical activity of
atomic gases. High precision of optical measurements allow us to expect that
investigation of time reversal invariance in photon interactions with atoms
will provide new limits for constants of $T$-noninvariant weak interaction.

As it was shown in \cite{Bar-in-E,Bar_lanl,Bar_magnet,Bar_PT} $T$
noninvariant interactions induce several new optical phenomena. They are: $T$
noninvariant rotation of polarization plane of light in electric field, $T$
noninvariant birefringence and $T$ noninvariant rotation of light
polarization plane in diffraction grating with non-centrosymmetrical
elementary cell.

To observe the rotation of polarization plane of light in electric
field experiments \cite{Budker,Bar-in-E} are under preparation now.
Therefore, it is important to attract attention to the fact that two
effects can contribute to $T$-noninvariant rotation of polarization
plane of light.  They are:

\begin{enumerate}
\item Light polarization plane rotation caused by the pseudo-Zeeman
  splitting of atomic levels in atoms with nonzero EDM in electric
  field \cite{Zeldovich,NMOE,Budker}.

\item Light polarization plane rotation due to interference of $P$-
  , $T$ - odd and Stark-induced transition amplitudes
  \cite{Bar-in-E,Bar_lanl,Bar_magnet}.
\end{enumerate}


The present paper is organized as follows. In Sec. II the general
theory describing $PT$ noninvariant rotation of polarization plane of
light in atomic gas is examined. Polarization plane rotation caused by
interference of $P$-, $T$- odd and Stark-induced transitions is
considered in details in Sec.  III. Polarization plane rotation caused
by nonzero atomic EDM is considered in Sec. IV. Estimates of effects
magnitude for different kinds of transitions are given in Sec. V. Sec.
VI examines possible sources of $P$-, $T$- violating interactions in
atom and Sec VII gives estimates of angle of $P$-, $T$- odd
polarization plane rotation for $6S_{1/2} \rightarrow 7S_{1/2}$
transition in cesium. Section VIII briefly discusses possibilities to
observe this phenomenon experimentally and summarizes general
conclusions.

\section{PT - odd rotation of polarization plane of light}

To illustrate the mechanism of polarization plane rotation due to
interference of $PT$ - odd and Stark-induced transition amplitudes we
consider a simple model at first. Let us take an atom in the $s_{1/2}$ state
and place it to an electric field. Taking into account the admixture of the
nearest $p_{1/2}$ state due to $P$- and $T$ - odd interactions and interaction
with electric field one can represent the wave function of an atom in the
form: 
\begin{eqnarray}
|{\ \widetilde s_{1/2} } \rangle &=& \frac{1}{\sqrt{4 \pi}} (R_0 (r) - R_1
(r) (\vec \sigma \vec n) \eta \\
& & - R_1(r) (\vec \sigma \vec n) (\vec \sigma \vec E) \delta ) |\chi_{1/2}
\rangle  \nonumber
\end{eqnarray}
Here $\vec \sigma$ are the Pauli matrices, $\vec n = \vec r /r$ is the unit
vector along $\vec r$, $\vec E$ is the electric field strength, $R_0$ and $%
R_1$ are radial parts of $s_{1/2}$ and $p_{1/2}$ wave functions
respectively, $| \chi_{1/2} \rangle$ is the spin part of wave function, $\eta
$ and $\delta$ are the mixing coefficients describing $P$ and $T$
noninvariant interactions and electric field respectively, $\vec E$ is the
electric field strength.

Interference of Stark and $PT$ - odd terms changes electron spin direction
as follows: 
\begin{eqnarray}
\Delta \vec s (\vec r) & = & \frac{ \eta \delta }{8 \pi} R_1^2 (r) \langle
\chi_{1/2} | (\vec \sigma \vec n) \vec \sigma (\vec \sigma \vec n) (\vec %
\sigma \vec E)  \nonumber \\
& & + (\vec \sigma \vec E) (\vec \sigma \vec n) \vec \sigma (\vec \sigma 
\vec n) | \chi_{1/2} \rangle  \nonumber \\
& = & \frac{\eta \delta R_1^2(r)}{ 8 \pi} \biggl( 4 \vec n (\vec n \vec E) -
2 \vec E \biggr)
\end{eqnarray}
Vector field $4 \vec n (\vec n \vec E) - 2 \vec E $ is shown in Fig. 1.
Since $\Delta \vec s$ does not depend on initial direction of atomic spin,
this spin structure appears even in nonpolarized atom. Spin vector averaged
over spatial variables differs from zero and is directed along $\vec E$.
Photons with angular moment parallel and antiparallel to $\vec E$
differently interact with such gas. It causes rotation of polarization
plane of photons.

Let us also note that according to \cite{Bar_magnet} magnetization of gas in
electric field induces magnetic field $\vec H_{ind}(E)$. This magnetic field
interacts with magnetic moment of an atom giving additional contribution to
rotation of polarization plane of light \cite{Bar_lanl}.

The refraction index of gas is given by 
\begin{equation}
n=1 + \frac{2 \pi N}{k^2} f(0)  \label{eq:basicBar}
\end{equation}
where $N$ is the number of atoms per $cm^3$, $k$ is the photon wave number, $%
f(0) = f_{ik} e_i^{*^{\prime}} e_k$ is the amplitude of elastic coherent
forward scattering of photons by atoms. Here $\vec e$ and $\vec e^{\prime}$
are the polarization vectors of initial and scattered photons respectively.
Repeated indices imply summation. In dipole approximation 
\begin{equation}
f_{ik} = \omega^2 \alpha_{ik} / c^2  \label{eq:pol_ampl}
\end{equation}
where $\alpha_{ik}$ is the tensor of dynamical polarizability of an atom, $%
\omega$ is the frequency of incident light. According to \cite%
{Bar-in-E,Kozlov2} the amplitude of light scattering by nonpolarized atomic
gas in electric field is expressed by 
\begin{eqnarray}
f_{ik} &=& f^{ev}_{ik}  \nonumber \\
& & + \frac{\omega^2}{c^2} (i \beta^P_s \epsilon_{ikl} n_{\gamma l} + i
\beta^{PT}_{E} \epsilon_{ikl} n_{El} + \beta^T_{sE} {(\vec n_{\gamma} \cdot 
\vec n_{E}) \delta_{ik}})  \label{eq:amplitude}
\end{eqnarray}
Here $f^{ev}_{ik}$ is the $P$- and $T$- invariant part of scattering
amplitude, $\beta^P_s$ is the $P$-odd but $T$-even scalar atomic
polarizability \cite{Khriplovich}, $\beta^{PT}_{E}$ is the $P$- and $T$- odd
scalar polarizability of an atom \cite{Bar-in-E}, $\beta^T_{sE}$ is the $P$%
-even but $T$- odd atomic polarizability \cite{Kozlov2}, $\vec n_{\gamma} = 
\vec k / k$ is the unit vector along the direction of photon propagation, $%
\vec n_{E} = \vec E / E$ is the unit vector along the direction of electric
field, $\epsilon_{ijk}$ is the third-rank antisymmetric tensor.

Angle of rotation of polarization plane is 
\begin{equation}
\phi = \frac{1}{2} k {\bf Re} (n_{+} - n_{-}) l  \label{eq:simple_angle}
\end{equation}
where $n_{+}$ and $n_{-}$ are the refraction indices for left and right
circularly polarized photons. Vectors $\vec e_{+}$ and $\vec e_{-}$ describe
left and right circularly polarized photons respectively, $\vec e_{\pm} =
\mp (\vec e_x \pm i \vec e_y)/ \sqrt{2}$. Using (\ref{eq:amplitude}) and (%
\ref{eq:basicBar}) we can express polarization plane rotation as follows 
\begin{equation}
\phi = - \frac{2 \pi N \omega}{c} (\beta^P_s + \beta^{PT}_E (\vec n_E \vec n%
_{\gamma})) l  \label{eq:angle}
\end{equation}
Term proportional to $\beta^P_s$ describes the well known phenomenon of $P$%
-odd but $T$-even rotation of polarization plane of light. Term proportional
to $\beta^{PT}_E$ corresponds to $P$- and $T$- noninvariant light
polarization plane rotation about the direction of electric field \cite%
{Bar_PT}.

In contrast to $P$-odd, $T$-even rotation the reversion of electric field
direction changes the sign of $P$-, $T$-odd rotation of light polarization
plane. This allows to distinguish $P$-, $T$-odd effects from the other possible
effects of polarization plane rotation.

According to \cite{Bar-in-E,Bar_lanl} tensor of dynamical polarizability of
an atom (molecule) in the ground state $|{\widetilde g_n}\rangle$ has the
form 
\begin{eqnarray}
\alpha_{ik}^{n} & = & \sum_m \Biggr\{ \frac{\left\langle {\widetilde g_n}
|d_i| {\widetilde e_m} \right\rangle \left\langle {\widetilde e_m} |d_k| {%
\widetilde g_n} \right\rangle }{ E_{em} - E_{gn} - \hbar \omega}  \nonumber
\\
& & + \frac{\left\langle {\widetilde g_n} |d_k| {\widetilde e_m}
\right\rangle \left\langle {\widetilde e_m} |d_i| {\widetilde g_n}
\right\rangle }{ E_{em} - E_{gn} + \hbar \omega} \Biggr\}
\label{eq:polarizability}
\end{eqnarray}
where $|{\widetilde g_n}\rangle$ and $|{\widetilde e_m}\rangle$ are the wave
functions of an atom in the ground and excited states perturbed by electric
field and $P$-, $T$- noninvariant interactions, $d$ is the operator of dipole
transition, $E_{em}$ and $E_{gn}$ are the energies of atom states $|{%
\widetilde g_n}\rangle$ and $|{\widetilde e_m}\rangle$, respectively.

In general case atoms are distributed over the magnetic sub-levels of ground
state $g_n$ with the probability $P(n)$. Therefore $\alpha_{ik}^{n}$ should
be averaged over all states $n$. As a result, the polarizability can be
written as 
\begin{equation}
\alpha_{ik} = \sum_{n} P(n) \alpha_{ik}^{n}  \label{eq:10-1}
\end{equation}
In the present paper we discuss the nonpolarized atomic gas. In this case $%
P(n) = 1/(2 j_g + 1)$ where $j_g$ is the total moment of atom in the ground
state $g$.

Tensor $\alpha_{ik}$ can be decomposed into irreducible parts as 
\begin{equation}
\alpha_{ik} = \alpha_{0}\delta_{ik} + \alpha_{ik}^{s} + \alpha_{ik}^{a}.
\label{eq:11}
\end{equation}
Here $\alpha _{0} = \frac{1}{3}{\ \sum_{i} \alpha _{ii}}$ is the scalar, $%
\alpha_{ik}^{s} = \frac{1}{2}(\alpha_{ik}+\alpha_{ki}) - \alpha_{0}
\delta_{ik}$ is the symmetric tensor of rank two, $\alpha_{ik}^{a}= \frac{1}{%
2}(\alpha_{ik}-\alpha_{ki})$ is the antisymmetric tensor of rank two: 
\begin{eqnarray}
\alpha_{ik}^a &=& \frac{\omega}{(2 j_g + 1)\hbar} \sum_{m,n}
\label{eq:antisym} \\
& & \Biggr\{ \frac{\left\langle {\widetilde g_n} |d_i| {\widetilde e_m}
\right\rangle \left\langle {\widetilde e_m} |d_k| {\widetilde g_n}
\right\rangle - \left\langle {\widetilde g_n} |d_k| {\widetilde e_m}
\right\rangle \left\langle {\widetilde e_m} |d_i| {\widetilde g_n}
\right\rangle }{ \omega^2_{em,gn} - \omega^2 } \Biggr\}  \nonumber
\end{eqnarray}
where $\omega_{em,gn}=(E_{em}-E_{gn}) / \hbar $.

If atoms are nonpolarized then in the absence of $P$- and $T$-odd
interactions the antisymmetric part of polarizability is equal to zero.
Therefore comparison of (\ref{eq:amplitude}) and (\ref{eq:pol_ampl}) yields 
\begin{equation}
\alpha^{a}_{ik} = i \epsilon_{ikl} (\beta^P_s n_{\gamma l} + \beta^{PT}_E
n_{El}).  \label{eq:alpha_beta}
\end{equation}
According to \cite{Bar_lanl,Bar_PT,Khriplovich} correct consideration of $P$%
-odd but $T$-even interactions requires to take into account both $E1$ and $%
M1$ transition amplitudes. If only $E1$ transition operators are considered
in (\ref{eq:polarizability}) the $P$- odd but $T$- even polarizability $%
\beta^P_s$ becomes equal to zero.

Evaluation of expression (\ref{eq:alpha_beta}) for left (or right) circular
polarization of incident light at $\vec n_{E} \parallel \vec n_{\gamma }$
yields $\alpha^{a}_{ik} e^{*(\pm)}_i e^{(\pm)}_k = \mp \beta^{PT}_E$. As
a result we can represent $P$-, $T$-odd scalar polarizability of an atom as
follows: 
\begin{eqnarray}
\beta_{E}^{PT} &= &\frac{\omega }{(2 j_g + 1)\hbar }\sum_{n,m} \Biggr\{ 
\frac{\left\langle {\widetilde g_n} |d_{-}| {\widetilde e_m} \right\rangle
\left\langle {\widetilde e_m} |d_{+}| {\widetilde g_n} \right\rangle } {
\omega^2_{em,gn} - \omega^2 }  \nonumber \\
& & - \frac{\left\langle {\widetilde g_n} |d_{+}| {\widetilde e_m}
\right\rangle \left\langle {\widetilde e_m} |d_{-}| {\widetilde g_n}
\right\rangle } { \omega^2_{em,gn} - \omega^2 } \Biggr\}  \label{eq:beta_pt}
\end{eqnarray}
where $d_{\pm} = \mp (d_x \pm i d_y)/ \sqrt{2}$.

For further analysis more detailed expressions for wave functions of an
atom are necessary. Since constants of $P$-, $T$- noninvariant interactions are
very small we can use perturbation theory. Let $|\overline g \rangle$ and $|%
\overline e \rangle$ be the wave functions of ground and excited states of
atom (molecule) in electric field $\vec E$ in the absence of $P$-, $T$- odd
interactions. Switch on $P$-, $T$- noninvariant interaction $(H_T\neq 0)$.
According to the perturbation theory the wave functions $|\widetilde g
\rangle$ and $|\widetilde e \rangle$ take the form 
\begin{eqnarray}
|{\widetilde g}\rangle & = & |\overline g\rangle + \sum_n |n\rangle \frac{%
\langle n |H_T| \overline g \rangle}{E_g - E_n}  \nonumber \\
|{\widetilde e}\rangle & = & |\overline e\rangle + \sum_n |n\rangle \frac{%
\langle n |H_T| \overline e \rangle}{E_e - E_n}  \label{eq:wave1}
\end{eqnarray}
where $H_T$ is Hamiltonian of $P$-, $T$- noninvariant interactions.

It should be reminded that denominator of (\ref{eq:beta_pt}) contains shifts
caused both by interaction of electric dipole moment of an atom with
electric field $\vec E$ and magnetic moment of an atom with $T$-odd induced
magnetic field $H_{ind}(\vec E)$ \cite{Bar_magnet}. If $H_T$ is small, one
can represent total polarizability $\beta_{E}^{PT}$ as the sum of two terms 
\begin{equation}
\beta_{E}^{PT}=\beta_{mix}+\beta_{split}.  \label{sum}
\end{equation}
Here 
\begin{eqnarray}
\beta_{mix} & = &\frac{\omega }{(2 j_g + 1)\hbar } \sum_{n,m} \Biggr\{ \frac{%
\left\langle {\widetilde g_n} |d_{-}| {\widetilde e_m} \right\rangle
\left\langle {\widetilde e_m} |d_{+}| {\widetilde g_n} \right\rangle} {
\omega^2_{\overline{e}m,\overline{g}n} - \omega^2 }  \nonumber \\
& & - \frac {\left\langle {\widetilde g_n} |d_{+}| {\widetilde e_m}
\right\rangle \left\langle {\widetilde e_m} |d_{-}| {\widetilde g_n}
\right\rangle } { \omega^2_{\overline{e}m,\overline{g}n} - \omega^2 } %
\Biggr\}  \label{eq:beta_mix}
\end{eqnarray}
where $\omega_{\overline{e}m,\overline{g}n} $ does not include $P$-, $T$-%
noninvariant shift of atomic levels, and 
\begin{eqnarray}
\beta_{split} &= &\frac{\omega }{(2 j_g + 1) \hbar }\sum_{n,m} \Biggr\{ 
\frac{\left\langle {\overline g_n} |d_{-}| {\overline e_m} \right\rangle
\left\langle {\overline e_m} |d_{+}| {\overline g_n} \right\rangle } {
\omega^2_{em,gn} - \omega^2 }  \nonumber \\
& & - \frac{\left\langle {\overline g_n} |d_{+}| {\overline e_m}
\right\rangle \left\langle {\overline e_m} |d_{-}| {\overline g_n}
\right\rangle } { \omega^2_{em,gn} - \omega^2 } \Biggr\}
\label{eq:beta_split}
\end{eqnarray}
\[
\omega_{em,gn} = ( E_{em}(\vec E) - E_{gn} (\vec E))/ \hbar 
\]
where energy levels $E_{e,m}(\vec E)$ and $E_{g,n}(\vec E)$ contain shifts
caused by interaction of electric dipole moment of an atom with electric
field $\overrightarrow{E}$ and magnetic moment of an atom with $T$-odd
induced magnetic field $\overrightarrow{H}_{ind}(\overrightarrow{E})$.

Below we consider small detuning of radiation frequency from resonance
frequency of atomic transition. Therefore (\ref{eq:beta_mix}) and (\ref%
{eq:beta_split}) can be written as follows 
\begin{eqnarray}
\beta_{mix} = \frac{1}{2 \hbar (2j_g+1)}\sum_{n,m} & & \Biggr\{ \frac{%
\left\langle {\widetilde g_n} |d_{-}| {\widetilde e_m} \right\rangle
\left\langle {\widetilde e_m} |d_{+}| {\widetilde g_n} \right\rangle } {
\omega_{\overline{e}m,\overline{g}n} - \omega }  \nonumber \\
& & - \frac {\left\langle {\widetilde g_n} |d_{+}| {\widetilde e_m}
\right\rangle \left\langle {\widetilde e_m} |d_{-}| {\widetilde g_n}
\right\rangle } { \omega_{\overline{e}m,\overline{g}n} - \omega } \Biggr\}
\label{eq:beta_mix1}
\end{eqnarray}
\begin{eqnarray}
\beta_{split}=\frac{1 }{2 \hbar (2 j_g + 1)} \sum_{n,m} & & \Biggr\{ \frac{%
\left\langle {\overline g_n} |d_{-}| {\overline e_m} \right\rangle
\left\langle {\overline e_m} |d_{+}| {\overline g_n} \right\rangle} {
\omega_{em,gn} - \omega }  \nonumber \\
& & - \frac{\left\langle {\overline g_n} |d_{+}| {\overline e_m}
\right\rangle \left\langle {\overline e_m} |d_{-}| {\overline g_n}
\right\rangle } { \omega_{em,gn} - \omega } \Biggr\}  \label{eq:beta_split1}
\end{eqnarray}

\section{Interference of PT - odd and Stark induced amplitudes}

In this section we consider effects associated with $\beta_{mix}$. Rotation
associated with $\beta_{split}$ is studied in Section IV.

Let us assume that electric field is small enough. For atoms in the ground
state the energy of Stark interactions is usually less than the difference
of energies of levels mixed by electric field. In this case we can use first
order of perturbation theory. Perturbed states $|{\widetilde g}\rangle$ and $%
|{\widetilde e}\rangle$ have the form 
\begin{eqnarray}
|{\widetilde g}\rangle = |g\rangle & + & \sum_n |n\rangle \frac{\langle n
|H_T| g \rangle}{E_g - E_n}  \nonumber \\
& + & \sum_m |m\rangle \frac{\langle m | - \vec d \vec E |g \rangle}{E_g -
E_m}  \nonumber \\
|{\widetilde e}\rangle = |e\rangle & + & \sum_n |n\rangle \frac{\langle n
|H_T| e \rangle}{E_e - E_n}  \nonumber \\
& + & \sum_n |m\rangle \frac{\langle m | - \vec d \vec E |e \rangle}{E_e -
E_m}.  \label{eq:eANDg_bar}
\end{eqnarray}
Here $H_T$ is the Hamiltonian of $P$-, $T$- noninvariant interactions, $| g
\rangle$ and $| e\rangle$ are the unperturbed ground and excited states of
an atom, $\vec E$ is the external electric field. We assume that electric
field is directed along $z$ axis.

Using (\ref{eq:eANDg_bar}) we can rewrite $\beta_{mix}$ as follows: 
\begin{eqnarray}
\beta_{mix} &=& \frac{1}{\hbar (2 j_g + 1)} {\rm Re}  \nonumber \\
& & \sum_{m_g, m_e} \frac {\left\langle g | d^{PT}_{+} | e \right\rangle
\left\langle e | d^{St}_{-} | g \right\rangle - \left\langle g | d^{PT}_{-}
| e \right\rangle \left\langle e | d^{St}_{+} | g \right\rangle } { \omega_{%
\overline{e}m,\overline{g}n} - \omega },  \label{eq:A_pm2}
\end{eqnarray}
where 
\begin{equation}
\left\langle g | d^{PT}_{\pm} | e \right\rangle = \sum_m \frac {\left\langle
g | H_T | m \right\rangle \left\langle m | d_{\pm} | e \right\rangle} {E_m -
E_g} + \frac {\left\langle g | d_{\pm} | m \right\rangle \left\langle m |
H_T | e \right\rangle} {E_m - E_e}  \label{eq:d_pt}
\end{equation}
and 
\begin{equation}
\langle g | \vec d^{St} \vec e | e \rangle = \Lambda_{ik} e_i E_k,
\label{eq:Stark0}
\end{equation}
is the Stark-induced amplitude of transition between states $g$ and $e$ in
the constant electric field $\vec E$, $\vec e$ is the polarization vector of
photon. 
\begin{equation}
\Lambda_{ik} = \sum_n \frac {\left\langle g | d_k | n \right\rangle
\left\langle n | d_{i} | e \right\rangle} {E_n - E_g} + \frac {\left\langle
g | d_{i} | n \right\rangle \left\langle n | d_{k} | e \right\rangle} {E_n -
E_e}.
\end{equation}
Representation of the second-rank tensors $\Lambda_{ik}$ and $e_i E_k$ in
terms of their irreducible spherical components yields \cite{Kozlov}: 
\begin{eqnarray}
\langle e | \vec d^{St} \vec e | g \rangle & = & \sum_{q,q^{\prime}}
(-1)^{q+q^{\prime}} \Lambda_{q,q^{\prime}} E_{-q} e_{-q^{\prime}}  \nonumber
\\
& = & \sum_{K,Q} (-1)^Q \Lambda^K_Q (E \otimes e)^K_{-Q},
\label{eq:Dst_lambda}
\end{eqnarray}
where subscripts $q$ and $q^{\prime}$ refer to the spherical vector
components, $\Lambda^K_Q$ and $(E \otimes e)^K_{-Q}$ are the components of
irreducible spherical tensors.

Using Wigner - Ekhard theorem we can represent $\Lambda^K_Q$ as follows: 
\begin{equation}
\Lambda^K_Q = (-1)^{j_e-m_e} \left( \matrix{ j_e & K & j_g \cr -m_e & Q &
m_g \cr} \right) \Lambda^K.  \label{eq:lambda}
\end{equation}
Reduced matrix elements $\Lambda^K$ ($K = 0, 1, 2$) are proportional to the
scalar, vector and tensor transition polarizability respectively.
Substituting (\ref{eq:Dst_lambda}), (\ref{eq:lambda}) into (\ref{eq:A_pm2}),
representing the matrix element of $PT$ - odd $E1$ transition (\ref{eq:d_pt}%
) in terms of reduced matrix element $\langle e || d^{PT} || g \rangle$ and
performing summation over $m_g$, $m_e$ one obtains the following expression
for $\beta_{mix}$ 
\begin{equation}
\beta_{mix} = - \frac{2}{3 \hbar (2 j_g + 1)} \frac{{\rm Re} \langle e ||
d^{PT} || g \rangle \Lambda^1 E } {\sqrt{2} (\omega_{\overline{e}m,\overline{%
g}n} - \omega )}  \label{eq:beta_summed}
\end{equation}
We assume here that electric field $\vec E$ is parallel to the direction of
light propagation and use the expression $(\vec E \otimes \vec e%
_{\pm})^1_{\pm} = E / \sqrt{2}$. Due to orthogonality of $3j$-symbols only
terms proportional to the vector part of transition polarizability remains
in (\ref{eq:beta_summed}) after summation over magnetic sub-levels.

Equations (\ref{eq:beta_summed}) and (\ref{eq:angle}) give the angle of
polarization plane rotation without considering of the Doppler broadening in
gas. Because of Doppler shift the resonance frequency of transition for a
single atom depends on atom velocity. In order to obtain expression for
angle of polarization plane rotation in this case we should average (\ref%
{eq:beta_summed}) over Maxwell distribution of atom velocity.

If nucleus has a nonzero spin we should take into account the hyperfine
structure. After routine calculations the angle of $P$- ,$T$-odd rotation of
polarization plane can be expressed as: 
\begin{eqnarray}
\phi &=& 4 \pi N_F l \frac{\omega}{\hbar c \Delta_D} g(u,v) \frac{1}{3 (2F_g
+ 1) } K^2  \nonumber \\
& & \times  {\rm Re} (\left\langle g || d^{PT} || e \right\rangle \Lambda^1
E \frac{1}{\sqrt{2}} ).  \label{eq:angle1}
\end{eqnarray}
For completeness we give here the expression for absorption length of light
in atomic gas \cite{Khriplovich} 
\begin{eqnarray}
L^{-1} & =& 4 \pi N_{F} \frac{\omega}{\hbar c \Delta_D} f(u,v) \frac{1}{3
(2F_g + 1) } K^2  \nonumber \\
& & \times | \left\langle g || A || e \right\rangle |^2  \label{eq:length1}
\end{eqnarray}
Here $F_g$, $F_e$ are the total angular moments of an atom in ground and
exited states respectively, $j_g$ and $j_e$ are the total electron moments
in these states, $i$ is the nuclear spin. 
\[
N_F =N \frac{2F_g+1}{(2i+1)(2j_g+1)}
\]
is the density of atoms with total moment $F_g$, 
\[
K^2 = (2 F_g+1) (2 F_e+1) \left\{ 
\matrix{
i  &  j_g &  F_g \cr
1  &  F_e &  J_e \cr} \right\} 
\]
$\Delta_{D}=\omega_0 \sqrt{2 k T / M c^2} $ is the Doppler line width,  
\begin{equation}
\begin{array}{l}
g(u,v) \\ 
f(u,v)
\end{array}
\Biggr\} = 
\begin{array}{l}
{\rm Im} \\ 
{\rm Re}%
\end{array}
\Biggr\} \sqrt{\pi }e^{-w^{2}}\left( 1-\Phi (-iw) \right),
\end{equation}
where $w=u+i v$, $\Phi(z) = \frac{2}{\sqrt{\pi}} \int^{z}_{0} dt e^{-t^2}$, $%
u=(\omega -\omega_0 )/\Delta_D $, $v=\Gamma / 2\Delta_{D} $, $\Gamma$ is the
recoil line width, $\left\langle g || A || e \right\rangle$ is the reduced
matrix element of dipole transition between states $| e \rangle$ and $| g
\rangle$.

\section{Rotation of polarization plane due to atomic EDM}

Presence of EDM in ground or excited state of atom also causes rotation of
polarization plane of light. We can derive the expression for the angle of
polarization plane rotation performing the calculations similar to those
described in Sec. III, but using $\beta_{split}$ instead of $\beta_{mix}$.
But in this case the calculations can be appreciably simplified if we note
that $P$-, $T$- noninvariant rotation caused by atomic EDM is similar to
Faraday rotation of photon polarization plane in a weak magnetic field.
Indeed, according to \cite{Khriplovich,Faraday} a weak magnetic field
affects the refractive index of atomic gas in two ways: through the change
of the magnetic sub-levels energies and through the mixing of hyperfine
states.

If we consider only terms proportional to the magnetic field strength $H$
and neglect the terms of higher orders then the level shift is \cite{Faraday}: 
\[
\Delta E_i = - H \langle i | \mu_z | i \rangle 
\]
The magnetic field $H$ mixes states of the same $F_z$ but different $F$, so
the state $| j \rangle$ becomes 
\[
| \overline{j} \rangle = | j \rangle - \sum_{k \not= j } H_z \frac{| k
\rangle \langle k | \mu_z | j \rangle} {E_k -E_j} 
\]
If atom has an EDM then electric field similarly affects the refraction
index (see (\ref{eq:beta_split})) and leads to atomic levels shift: 
\[
\Delta E_i = - E \langle i | d_z | i \rangle 
\]
It also mixes the hyperfine states of atom with the same $F_z$ but different $F$ 
\[
| \overline j \rangle = | j \rangle - \sum_{k \not= j } E_z \frac{| k
\rangle \langle k | d_z | j \rangle} {E_k -E_j} 
\]
As a result we can use the expression for rotation of polarization plane of
light in a weak magnetic field \cite{Khriplovich,Faraday} for calculations of
effect of polarization plane rotation in electric field. For this we must
substitute $H \rightarrow E$, $\mu_g \rightarrow d_g$, $\mu_e \rightarrow d_e
$ where $d_e$, $d_g$ are EDM of atom in ground and excited states, $\mu_i$
is the magnetic moment of state $i$. 

If we neglect quadrupole transition amplitudes (it is possible for example
for $6S_{1/2} \rightarrow 7S_{1/2}$ transition in cesium), then the angle of
polarization plane rotation has the form 
\begin{eqnarray}
\phi & = & \frac{2 \pi N l}{(2 i+1)(2 j_g +1)} \frac{\omega}{\Delta_D \hbar c%
} \frac{E_z}{\hbar \Delta_D} |\left\langle g || A || e \right\rangle|^2 
\nonumber \\
& & \times \biggl( \frac{\partial g(u,v)}{\partial u} \delta_1 + 2 g(u,v)
\gamma_1 \biggr).  \label{eq:dipole}
\end{eqnarray}
The expressions for parameters $\gamma_1$ and $\delta_1$ are given below %
\onecolumn 
\begin{eqnarray}
\gamma_1 &=& \frac{(2 F_g + 1)(2 F_e + 1)}{\sqrt{6}} (-1)^i \left\{ \matrix{
i & j_g & F_g \cr 1 & F_e & j_e \cr} \right\} [ d_e (-1)^{j_e + F_g} \sqrt{%
\frac{(j_e + 1)(2 j_e +1)}{j_e}}  \nonumber \\
& & ( \frac{\Delta_D}{\Delta_{hf}(F_e, F_e -1)} (2 F_e -1) \left\{ \matrix{
i & j_g & F_g \cr 1 & F_e -1 & j_e \cr} \right\} \left\{ \matrix{ i & j_e &
F_e \cr 1 & F_e -1 & j_e \cr} \right\} \left\{ \matrix{ F_g & 1 & F_e \cr 1
& F_e -1 & 1 \cr} \right\}   \nonumber \\
& & + \frac{\Delta_D}{\Delta_{hf}(F_e, F_e +1)} (2 F_e +3) \left\{ \matrix{
i & j_g & F_g \cr 1 & F_e +1 & j_e \cr} \right\} \left\{ \matrix{ i & j_e &
F_e \cr 1 & F_e +1 & j_e \cr} \right\} \left\{ \matrix{ F_g & 1 & F_e \cr 1
& F_e +1 & 1 \cr} \right\} )   \nonumber \\
& & - \biggl( j_e \leftrightarrow j_g, F_e \leftrightarrow F_g, d_e
\leftrightarrow d_g, \biggr) ]  \nonumber
\end{eqnarray}
and 
\begin{eqnarray}
\delta_1 &=& \frac{(2 F_g + 1)(2 F_e + 1)}{\sqrt{6}} (-1)^i \left\{ \matrix{
i & j_g & F_g \cr 1 & F_e & j_e \cr} \right\}^2 [ d_e (-1)^{j_e + F_g} \sqrt{%
\frac{(j_e + 1)(2 j_e +1)}{j_e}}  \nonumber \\
& & (2 F_e +1) \left\{ \matrix{ i & j_e & F_e \cr 1 & F_e & j_e \cr}
\right\} \left\{ \matrix{ F_g & F_e & 1 \cr 1 & 1 & F_e \cr} \right\} + %
\biggl( j_e \leftrightarrow j_g, F_e \leftrightarrow F_g, d_e
\leftrightarrow d_g, \biggr) ]  \nonumber
\end{eqnarray}
\twocolumn  Here $\Delta_{hf}$ is the hyperfine level splitting. First term
in (\ref{eq:dipole}) arises from the level splitting in electric field. It
describes effect similar to Macaluso - Corbino rotation of photon
polarization plane in magnetic field. Second term is caused by mixing of
hyperfine levels with the different total moment $F$ but the same $F_z$ in
electric field. It describes the $T$ noninvariant analog of polarization
plane rotation due to Van-Vleck mechanism.

\section{Estimates}

Let us compare the angle of $P-,T$-odd polarization plane rotation for
different kinds of transitions. Angle of rotation of polarization plane per
absorption length due to interference of $P-,T$-odd and Stark - induced
transition amplitudes according to (\ref{eq:angle1}) is expressed by 
\begin{equation}
\phi(L_{abs}) = \frac{g(u,v)}{f(u,v)} \frac{{\rm Re} \left\langle g ||
d^{PT} || e \right\rangle \Lambda^1 E} {\sqrt{2} | \left\langle g || A || e
\right\rangle |^2}  \label{eq:estimate1}
\end{equation}
If detuning $\Delta \sim \Delta_D$ then $g \sim f \sim 1$.

The value of transition matrix element depends on kind of transition. For
allowed $E1$ transition $\left\langle g || A || e \right\rangle \sim
\left\langle d \right\rangle \sim e a_0$, for allowed $M1$ transition $%
\left\langle g || A || e \right\rangle \sim \left\langle  \mu \right\rangle
\sim \alpha \left\langle d \right\rangle$. 
For strongly forbidden $M1$ transition in electric field dominant
contribution 
to the angle of rotation 
gives the Stark-induced $E1$ transition. 
Its amplitude can be estimated as $\left\langle g || A || e \right\rangle
\sim \left\langle d \right\rangle^2 E_z / \Delta E$ where $\Delta E$ is the
typical difference between the opposite parity levels in atom. For
transition $6S_{1/2} \rightarrow 7S_{1/2}$ in $Cs$ and experimentally
accessible strength of electric field $E \sim 10^{3} - 10^{4}$ V/cm, value
of $\left\langle d \right\rangle E_z / \Delta E \sim 10^{-3} - 10^{-4}$.
Here $E_z$ is the electric field strength, $e$ is the electron charge, $a_0$
is the Bohr radius, $\Delta E \sim Ry$ is the typical difference between the
energy levels of the opposite parity states, $Ry = 13.6 {\rm eV}$ is the
Rydberg energy constant, $\alpha \simeq 1/137$ is the fine structure
constant.

The numerator of (\ref{eq:estimate1}) has the same order of magnitude for
all kinds of transitions considered above: $\left\langle g || d^{PT} || e
\right\rangle \Lambda^1 E \sim \left\langle d \right\rangle^2 \left\langle
H_T \right\rangle \left\langle d \right\rangle E_z / (\Delta E)^2 $. Now we
can estimate the angle of polarization plane rotation per absorption length.

For allowed $E1$ transition 
\begin{equation}
\phi( L_{abs} ) \sim \frac {\left\langle H_T \right\rangle} { \Delta E} 
\frac{\left\langle d \right\rangle E_z }{\Delta E},  \label{eq:E1_allowed}
\end{equation}
where $\left\langle H_T \right\rangle$ is the typical value of matrix
element of $PT$ - odd Hamiltonian.

The angle of rotation per absorption length near allowed $M1$ transition is
larger than in $E1$ case 
\begin{equation}
\phi( L_{abs} ) \sim \frac {\left\langle H_T \right\rangle} { \alpha^2
\Delta E} \frac{\left\langle d \right\rangle E_z }{\Delta E}.
\label{eq:M1_allowed}
\end{equation}
The largest value of angle of rotation per absorption length can be observed
near strongly forbidden $M1$ transition because the absorption of light is
the lowest in this case 
\begin{equation}
\phi( L_{abs} ) \sim \frac {\left\langle H_T \right\rangle} { \Delta E} 
\frac{\Delta E }{\left\langle d \right\rangle E_z }.  \label{eq:M1_forbidden}
\end{equation}

It is interesting to compare these estimates with angle of rotation of
polarization plane caused by nonzero EDM of atom. In the absence of
hyperfine structure the angle of rotation per absorption length can be
estimated using (\ref{eq:angle1}) and (\ref{eq:dipole}) as follows 
\begin{eqnarray}
\phi_{EDM}(L_{abs}) & = & \frac{1}{2(2 j_g +1)} \frac{E_z \delta}{f \Delta_D}
\frac{\partial g}{\partial u}  \nonumber \\
& \sim & \frac{d_{at} E_z}{\Delta_D} \sim \frac{\left\langle d \right\rangle
E_z}{\Delta_D} \frac{\left\langle H_T \right\rangle}{\Delta E},
\label{eq:phi_edm}
\end{eqnarray}
where $d_{at} \sim \langle d \rangle \langle H_T \rangle / \Delta E$ is the
EDM of atom, $\Delta_D \sim (10^{-5} \sim 10^{-6}) \Delta E$ is the Doppler
line width. The value $\phi (L_{abs})$ here does not depend on transition
amplitude $\left\langle g || A || e \right\rangle$ and has the same order of
magnitude for all kinds of transition considered above.

For allowed $E1$ transition $\phi(L_{abs}) / \phi_{EDM}(L_{abs}) \sim
\Delta_D / \Delta E << 1$ and dominant contribution to the angle of $P$-, $T$%
-odd polarization plane rotation gives the level splitting caused by atomic
EDM.

Near allowed $M1$ transition $\phi(L_{abs}) / \phi_{EDM}(L_{abs}) \sim
\Delta_D / \alpha^2 \Delta E \sim 1$ and both mechanisms contribute
comparably.

Near the strongly forbidden $M1$ transition $\phi(L_{abs}) /
\phi_{EDM}(L_{abs}) \sim \Delta_D \Delta E/ (\left\langle d \right\rangle
E_z)^2 >> 1$ and the interference of $P-,T$-odd and Stark-induced transition
contributes the most to the angle of polarization plane rotation.

\section{P and T - odd interactions in atom}

Several mechanisms can induce the violation of $P$- and $T$- invariance in
atom. According to \cite{Khriplovich} they are: $P$-, $T$- odd weak interactions
of electron and nucleon, interaction of electric dipole moment of electron
with electric field inside atom, interaction of electrons with electric
dipole and magnetic quadrupole moments of nucleus and $P$-, $T$-odd
electron-electron interaction.

We consider effects that according to \cite{Khriplovich} give the dominant
contribution in our case: $P$-, $T$-odd electron-nucleon interaction and
interaction of electron EDM with electric field inside atom.

According to \cite{Khriplovich,Barr,Hunter} Hamiltonian of $T$-violating
interaction between electron and hadron has the form: 
\begin{equation}
H_T = C_s \frac{G}{\sqrt{2}}(\bar e i \gamma_5 e) (\bar n n) + C_t \frac{G}{%
\sqrt{2}} (\bar e i \gamma_5 \sigma_{\mu \nu} e) (\bar n \sigma^{\mu \nu} n)
\label{eq:hep_hamiltonian}
\end{equation}
where $G=1.055 \cdot 10^{-5} m_p^{-2} $ is the Fermi constant, $m_p$ is the
proton mass, $e$ and $n$ are the electron and hadron field operators
respectively, $C_s$ and $C_t$ are the dimensionless constants characterizing
the strength of $T$-violating interactions relatively to $T$-conserving weak
interaction. The first term in (\ref{eq:hep_hamiltonian}) describes scalar
hadronic current coupling with pseudoscalar electronic current, and the
second one describes tensor hadronic current coupling with the pseudotensor
electronic current.

Matrix element of $T$ - odd Hamiltonian according to \cite{Khriplovich} is
equal to 
\begin{equation}
\left\langle s_{1/2} || H_T || p_{1/2} \right\rangle = \frac{G m_e^2
\alpha^2 Z^2 R}{2 \sqrt{2} \pi} \frac{{\bf Ry}}{\sqrt{ \nu_s \nu_p}^3} 2
\gamma C_s A  \label{Todd}
\end{equation}
where $m_e$ is the electron mass, $\nu_i$ is the effective principal quantum
number of state $i$, $A$ is the atomic number, $\gamma = \sqrt{ (j+1/2)^2
-Z^2 \alpha^2 }$ and $j$ is the total angular moment of atom. The
``relativistic enhancement factor'' $R$ is given by 
\[
R = 4 \frac{(a_0/2 Z r_0)^{2 - 2 \gamma}}{\Gamma^2 (2 \gamma + 1)}. 
\]
Here $r_0 = A^{1/3} 1.2 \cdot 10^{-13} {\rm cm}$ is the approximate nuclear
radius. For cesium $R=2.8$. We neglect tensor part of interaction for
simplicity.

Hamiltonian of interaction of electron EDM and electric field inside atom
that mix opposite parity atomic states has the form \cite{Khriplovich} 
\begin{equation}
H_d = \sum_k (\gamma_{0k} - 1) \vec \Sigma_k \vec E_k  \label{eq:EDM1}
\end{equation}
where 
\begin{equation}
\Sigma_k = - \gamma_5 \gamma_0 \gamma_k = \left( \matrix{ \sigma_k & 0 \cr 0
& \sigma_k \cr} \right),
\end{equation}
$\sigma_k$ are the Pauli matrices, $E_k$ is the electric field strength
acting upon electron $k$. If summation in (\ref{eq:EDM1}) is performed over
one valence electron and electric field strength near the nucleus
approximately equals $\vec E=Z \alpha \vec r / r^3$, matrix element of
operator $H_d$ can be written as follows \cite{Khriplovich} 
\begin{eqnarray}
\langle j, l=j+1/2 || & H_d &|| j, l^{\prime}=j-1/2 \rangle  \nonumber \\
& & = - \frac{4 (Z \alpha)^3 }{\gamma (4 \gamma^2 -1) (\nu_l
\nu_{l^{\prime}})^{3/2} a_0^2}  \label{eq:EDM_mat_el}
\end{eqnarray}
where $l$ and $l^{\prime}$ are the orbital angular moments.

\section{Estimates for $6s_{1/2} \rightarrow 7s_{1/2}$ cesium transition}

Let us estimate $P$-, $T$-odd rotation of polarization plane for highly
forbidden $M1$ transition $6s_{1/2} \rightarrow 7s_{1/2}$ in cesium. The
scheme of cesium energy levels is shown in Fig. 2.

\subsection{Rotation of polarization plane of light due to electron nucleon
interactions}

The $P$ and $T$ odd electron nucleon interactions mix $s$ and $p$ states of
cesium. Since the $E1$ transition amplitudes $6s \rightarrow np$ and $7s
\rightarrow n^{\prime}p$ are negligibly small when $n > 6$ and $n^{\prime} >
7$ \cite{Khriplovich} we should take into account only admixture of $6p_{1/2}
$ and $7p_{1/2}$ states. Using Eq. (\ref{eq:wave1}) and matrix element (\ref%
{Todd}) one can represent the wave functions perturbed by $P$ and $T$
noninvariant electron-nucleon interaction as follows: 
\begin{eqnarray}
|\widetilde{ 6s_{1/2}} \rangle = |6s_{1/2} \rangle & + & 10^{-11} \left( 2
\gamma \frac{A}{N} C_s \right)  \nonumber \\
& & \times (1.17 |6p_{1/2} \rangle + 0.34 |7p_{1/2} \rangle)  \nonumber \\
|\widetilde{ 7s_{1/2}} \rangle = |7s_{1/2} \rangle & + & 10^{-11} \left( 2
\gamma \frac{A}{N} C_s \right)  \nonumber \\
& & \times (0.87 |6p_{1/2} \rangle - 1.33 |7p_{1/2} \rangle ),
\label{Cs_state}
\end{eqnarray}
Where $N$ is the number of neutrons in atomic nucleus. Using Eq. (\ref%
{eq:wave1}) and values of radial integrals \cite{Khriplovich} $%
\rho(6s_{1/2},6p_{1/2})= -5.535$, $\rho(7s_{1/2},6p_{1/2})= 5.45$, $%
\rho(7s_{1/2},7p_{1/2})= -12.30$ we obtain reduced matrix element of $PT$ -
odd $E1$ transition 
\begin{equation}
\left\langle 6 s_{1/2} || d^{PT} ||7 s_{1/2} \right\rangle = 1.27 \cdot
10^{-10} |e| a_0 C_s
\end{equation}

The matrix element of Stark-induced $6s_{1/2} \rightarrow 7s_{1/2}$
transition in cesium is usually written as \cite{Bouchiat}: 
\[
\langle 6s_{1/2}, m^{\prime}| d_i^{St} | 7s_{1/2}, m \rangle = \alpha E_i
\delta_{mm^{\prime}} + i \beta \epsilon_{ijk} E_j \langle m^{\prime}|
\sigma_k | m \rangle 
\]
where $m$ and $m^{\prime}$ are the magnetic quantum numbers of ground and
excited states of cesium, $E_i$ is the electric field strength, $\alpha$ and 
$\beta$ are the scalar and vector transition polarizability (see also (\ref%
{eq:Dst_lambda})). The value of $\Lambda^1$ introduced in (\ref{eq:angle1})
can be expressed for cesium via the vector transition polarizability as
follows: $\Lambda_1 = - 2 \sqrt{3} \beta E$. Value of $\beta$ is well known
from theoretical calculations \cite{beta} as well as from experiment \cite%
{beta2}. According to \cite{beta} it is equal to $\beta = 27.0 a^3_0$.
Therefore 
\begin{equation}
\Lambda^1 = - 1.81 \cdot 10^{-8} |e| a_0 E(V/cm)
\end{equation}

Suppose temperature is $T=750 K$. Then pressure of $Cs$ vapor is $p=10$ kPa %
\cite{BigBlue}, concentration of atoms is $N= 10^{18} cm^{-3}$ and Doppler
line width is $\Delta_D / \omega_0 = 10^{-6}$.

For transition between hyperfine levels $F_g = 4 \rightarrow F_e = 4$
coefficient $K^2$ in formula (\ref{eq:angle1}) is maximal ($K^2=15/8$).

Suppose detuning $\Delta \sim \Delta_D$, then $v = \Gamma / 2 \Delta_D
\simeq 0.1 $ and $f \approx 1$, $g \approx 0.7$. Absorption length in
longitudinal electric field $E = 10^4 {\rm V/cm}$ is equal to $L_{abs}= 7 
{\rm m} $.

As a result angle of $PT$ noninvariant rotation of polarization plane is 
\[
|\phi| = 1.0 \cdot 10^{-13} C_s l E .
\]
The optimal signal to noise ratio is achieved when $l = 2 L_{abs}$ \cite%
{Khriplovich}. The best limit on the parameters of electron-nucleon
interaction $C_s < 4 \cdot 10^{-7}$ was set in \cite{Tl-EDM}. Corresponding
limit to the angle of rotation of polarization plane is $|\phi| < 0.5 \cdot
10^{-12} \;{\rm rad.}$

\subsection{Rotation of polarization plane of light due to cesium EDM}

Using wave functions (\ref{Cs_state}) we can obtain EDM in $6s_{1/2}$ and $%
7s_{1/2}$ states of $Cs$ 
\[
d_{6s_{1/2}} = - 1.35 \cdot 10^{-10} C_s |e| a_0
\]
\[
d_{7s_{1/2}} = - 4.39 \cdot 10^{-10} C_s |e| a_0 .
\]
As a result, expression (\ref{eq:dipole}) yields the angle of
polarization plane rotation due to level splitting in electric field 
\[
|\phi_1| = 1.4 \cdot 10^{-24} l C_s E^3_z ({\rm V/cm}) < 8 \cdot 10^{-16} 
\]
and the angle of rotation due to hyperfine levels mixing 
\[
|\phi_2| = 2.1 \cdot 10^{-24} l C_s E^3_z ({\rm V/cm}) < 1.2 \cdot 10^{-15}.
\]
(We assume here that for detuning $\Delta \sim \Delta_D$ functions
$g(u,v) \simeq 0.7$, $\partial g(u,v)/ \partial u \simeq 1.1$). These
angles are three orders of magnitude lower than angle of polarization
plane rotation arising from interference of Stark-induced and
$P$-, $T$- noninvariant transition amplitudes.

\subsection{Rotation of polarization plane of light due to electron EDM}

If the $PT$ noninvariant interaction in atom is induced by interaction of
electron EDM with the electric field of nucleus then using (\ref{eq:wave1})
and (\ref{eq:EDM_mat_el}) one can represent the wave functions of $6s$ and $%
7s$ states of cesium as follows 
\begin{eqnarray}
|{\widetilde 6s_{1/2}} \rangle & = & |{6s_{1/2}}\rangle  \nonumber \\
& & - (35 |6p_{1/2}\rangle + 10.5 |7p_{1/2}\rangle) d_e/ (e a_0)  \nonumber
\\
|{\widetilde 7s_{1/2}}\rangle & = & |{7s_{1/2}}\rangle  \nonumber \\
& & + (27.7 |6p_{1/2}\rangle - 36.2 | 7p_{1/2}\rangle d_e/ (e a_0 )
\label{eq:EDM2}
\end{eqnarray}
Using Eq. (\ref{eq:EDM2}) we can obtain reduced matrix element of electric
dipole transitions between ${\widetilde 6s_{1/2}}$ and ${\widetilde 7s_{1/2}}
$ states 
\begin{equation}
\left\langle {\ 6s_{1/2}} || d^{PT} || {\ 7s_{1/2}} \right\rangle = - 72 d_e
\label{eq:matel_electron}
\end{equation}
and electric dipole moment of cesium in ground state $d_{6s_{1/2}}$ and
excited state $d_{7s_{1/2}}$ 
\begin{eqnarray}
d_{6s_{1/2}} &=& 131 d_e  \nonumber \\
d_{7s_{1/2}} &=& 400 d_e ,  \label{eq:EDM_electron}
\end{eqnarray}
where $d_e$ is the electron EDM.

As we mentioned before, two effects induce $T$ - noninvariant rotation of
polarization plane in electric field. The first of them is the interference
of $PT$ - odd and Stark induced transition amplitudes and the second is the
interaction of atomic EDM with electric field. After substitution of (\ref%
{eq:matel_electron}) to (\ref{eq:angle1}) one can obtain the angle of
rotation arising from interference of amplitudes $|\phi| < 0.6 \cdot 10^{-12}
$ in the same experimental conditions as before.

Rotation caused by atomic EDM is a sum of two contributions. Using the first
term in (\ref{eq:dipole}) and (\ref{eq:EDM_electron}) one can obtain the
angle of rotation induced by splitting of magnetic sub-levels in electric
field $|\phi_1| < 1.3 \cdot 10^{-15}$. The mix of hyperfine components
(second term in Eq. (\ref{eq:dipole})) gives the contribution $|\phi_2| < 2
\cdot 10^{-15}$.

For estimates we use experimental limit on electron EDM from \cite{Tl-EDM} 
$|d_e| < 4 \cdot 10^{-27} |e| \; {\rm cm}$. We should note that estimates
of rotation angle for electron-nucleon $T-$ noninvariant weak interactions
and electron EDM give close values.

\section{Conclusion}

In the present article we have considered phenomenon of rotation of
polarization plane of light in gas placed in electric field. Calculations of
angle of polarization plane rotation are performed for $6S_{1/2} \rightarrow
7S_{1/2}$ transition in atomic cesium. Two mechanisms of effect are
considered. They are: 

\begin{enumerate}
\item Light polarization plane rotation caused by the  pseudo-Zeeman
splitting of atomic levels in atoms with nonzero  EDM in electric field \cite%
{Zeldovich,NMOE,Budker}.

\item Light polarization plane rotation due to interference of $PT$  - odd
and Stark-induced transition amplitudes \cite{Bar-in-E,Bar_lanl,Bar_magnet}.
\end{enumerate}


Both of them can be induced by $PT$ noninvariant interaction between
electrons and atomic nucleus and by interaction of electron EDM with
electric field inside atom.

For the highly forbidden $M1$ transition $6S_{1/2} \rightarrow 7S_{1/2}$ in
cesium we can expect the angle of polarization plane rotation per absorption
length due to $PT$ - odd atomic polarizability $\beta^{PT}_{mix}$ about $%
|\phi| < 10^{-12}$. The rotation induced by atomic EDM for this transition
is three orders of magnitude smaller.

Angle of polarization plane rotation can be significantly greater for other
atoms, for example, for rare-earth elements, where additional amplification
arises from close levels of opposite parity.

The simplest experimental scheme to observe pseudo-Faraday rotation of
polarization plane of light in electric field includes a cell with atomic
gas placed in the electric field and sensitive polarimeter. In case of large
absorption length one can place this cell in resonator or delay line optical
cavity to reduce the size of experimental setup (see e. q. \cite{Axions}).

Several schemes are proposed to increase the sensitivity of measurements.
One of them is based on the nonlinear magneto-optic effect (NMOE) \cite%
{NMOE,Budker}. Since the change of rotation angle with the change of applied
field in this case is several orders of magnitude larger than in the
traditional scheme, sensitivity of this kind of experiment can be very
high. The authors of \cite{Budker} expect to achieve the sensitivity for the
cesium EDM $|d_{Cs}| < 10^{-26} |e| $ cm. The corresponding limit for the
electron EDM is $|d_{e}| < 10^{-28} |e| $ cm.

The method of measurement of polarization plane rotation proposed in \cite%
{Bar-in-E,Bar_lanl,Bar_magnet} probably can provide even higher sensitivity.
This method is based on observation of evolution of polarization of light in
a a cell with atomic vapor and amplifying media placed in a resonator.
According to \cite{Bar-in-E,Bar_lanl,Bar_magnet} the compensation of
absorption of light in a cell allows to increase the observed angle of
polarization plane rotation and, according to estimates \cite{Bar_lanl},
allows to increase the sensitivity for electron EDM up to $|d_{e}| <
10^{-30} |e| $ cm.

Therefore we can hope that experimental measurement of described phenomenon
can provide sensitivity for parameters of $P-,T-$ noninvariant interactions
between electron and nucleus and electron EDM, comparable or even higher
than current atomic EDM experiments.

\begin{figure}
\epsfxsize =8.5cm
\centerline{\epsfbox{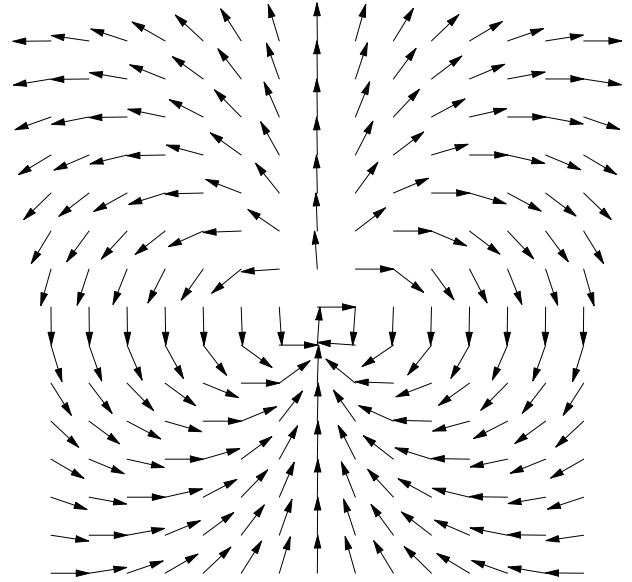}}
\vspace{7mm}
\caption{\it Vector field $4 \vec n (\vec n \vec E) - 2 \vec E$.
Vectors on figure shows direction of atomic spin in $s_{1/2}$ state if we take into
account admixture of $p_{1/2}$ state due to $PT$ noninvariant interactions
and external electric field.}
\vspace{0.5cm}
\label{figb}
\end{figure}

\begin{figure}
\epsfxsize =6.5cm
\centerline{\epsfbox{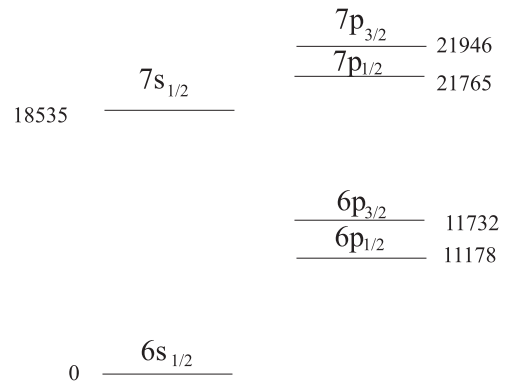}}
\vspace{7mm}
\caption{\it
Scheme of cesium energy levels. Energy of atomic levels
is given in $cm^{-1}$. }
\vspace{0.5cm}
\label{figc}
\end{figure}

\end{document}